\documentclass[12pt]{article}
\usepackage{makeidx}
\usepackage{amsfonts}
\usepackage{amsmath}
\usepackage{amssymb}
\usepackage[T1]{fontenc}

\usepackage{epsfig}
\usepackage{enumitem}
\usepackage{float}

\DeclareMathOperator\Tr{Tr\, }
\newtheorem{theorem}{Theorem}

\newtheorem{result}[theorem]{Result}

\numberwithin{equation}{section} \numberwithin{theorem}{section}
\begin{document}

\title{Entanglement Entropy of Free Fermions with a Random Matrix as a One-Body Hamiltonian}
\author{L. Pastur$^{1,2}$ and V. Slavin$^{1}$\\
$^{1)}$ B. Verkin Institute for Low Temperature Physics and Engineering\\ Kharkiv, Ukraine\\
$^{2)}$ King's College, London,UK}
\date{}
\maketitle

\begin{abstract}
We consider a quantum system
of large size $N$ and its subsystem of size $L$ assuming that $N$ is much larger than $L$, which can also be sufficiently large, i.e., $1 \ll L \lesssim
N $. A widely accepted mathematical version of this heuristic inequality
is the asymptotic regime of successive limits: first the macroscopic limit $%
N \to \infty$, then an asymptotic analysis of the entanglement entropy as
$L \to \infty$. In this paper, we consider another version of the above
heuristic inequality: the regime of asymptotically proportional $L$ and $N$, i.e., the simultaneous limits $L \to \infty,\; N \to \infty, L/N \to \lambda >0$. Specifically, we consider the system of free fermions which is in its ground state and such that its one-body Hamiltonian is a large random matrix, that is often used to model the long-range hopping. By using random matrix theory, we
show that in this case, the entanglement entropy obeys the volume law known for
systems with short-ranged hopping but described either by a mixed
state or a pure strongly excited state of the Hamiltonian. We also give a streamlined proof
of Page's formula for the entanglement entropy of the
black hole radiation for a wide class of typical ground states, thereby proving the universality of the formula.
\end{abstract}


\section{Introduction}

\label{s:int}

Quantum entanglement, a special form of quantum correlation, is regarded as
one of the important ingredients of modern quantum mechanics and adjacent
fields of science and technology. 
In its simplest form, the entanglement causes two quantum objects (spins, qubits, etc.)
to share a common pure state in which they do not have pure states of their
own.


A general version of this simplest form is known as the bipartite setting  where a quantum system $\mathfrak{S}$ consists of  parties $%
\mathfrak{B}$ and $\mathfrak{E}$, i.e., symbolically
\begin{equation}
\mathfrak{S}=\mathfrak{B}\cup \mathfrak{E}.  \label{SBE}
\end{equation}%
Sometimes the parties are two communicating agents, sometimes one of them,
say $\mathfrak{B}$ (block), is the system of interest while $\mathfrak{E}$
is the environment of $\mathfrak{B}$, etc. There is a variety of versions
and models for this general setting and related problems, see e.g. \cite%
{Ab-Co:19,Al-Co:21,Am-Co:08,Ao-Co:15,Ca-Co:11,
Da-Co:14,De:18,Ho-Co:09,Pa:05,Re-Mo:09,Sz-Co:20,Wi:18}
for reviews.

Denote by $\mathcal{H}_{\mathfrak{S}}$, $\mathcal{H}_{\mathfrak{B}}$, and $%
\mathcal{H}_{\mathfrak{E}}$ the corresponding state spaces, so that
\begin{equation}
\mathcal{H}_{\mathfrak{S}}=\mathcal{H}_{\mathfrak{B}}\otimes \mathcal{H}_{%
\mathfrak{E}},  \label{stsp}
\end{equation}%
and by $\mathrm{tr}_{\mathfrak{B}}$ and $\mathrm{tr}_{\mathfrak{E}}$ the
operation of (partial) traces    in $\mathcal{H}_{\mathfrak{B}}$ and $\mathcal{H%
}_{\mathfrak{E}}$. 
Let $\rho _{\mathfrak{S}}$ be the density matrix of $\mathfrak{S}$, which is
often assumed to be in a pure state, i.e.,
\begin{equation}
\rho _{\mathfrak{S}}=|\Psi _{\mathfrak{S}}\rangle \langle \Psi _{\mathfrak{S}%
}|.  \label{gss}
\end{equation}%
Applying $\mathrm{tr}_{\mathfrak{E}}$ to $\rho _{\mathfrak{S}}$, we obtain
the reduced density matrix
\begin{equation}
\rho _{\mathfrak{B}\mathfrak{S}}=\mathrm{tr}_{\mathfrak{E}}\,\rho _{%
\mathfrak{S}},\;\mathrm{tr}_{\mathfrak{B}}\,\rho _{\mathfrak{B}\mathfrak{S}%
}=1  \label{rdmg}
\end{equation}%
of the $\mathfrak{B}$, a positive definite operator acting in $\mathcal{H}_{%
\mathfrak{B}}$. It can be viewed as quantum analog of the marginal
distribution of probability theory.

If $\mathfrak{B}$ consists of several (1, 2, etc.) elementary objects, then the
corresponding reduced density matrices are known in quantum statistical
mechanics as the one-, two-, etc.-point correlation functions. In this paper
we will deal with extended systems and their subsystems (parties), hence,
with reduced density matrices (correlation functions) of large size.

One of widely used numerical characteristics (quantifiers) of  the quantum correlations between the parties is the  entanglement is entropy 
\begin{equation}
S_{\mathfrak{B}\mathfrak{S}}=-\mathrm{tr}_{\mathfrak{B}}\rho _{\mathfrak{B}}\log _{2}\rho _{\mathfrak{B}\mathfrak{S}},
\label{ente}
\end{equation}%
i.e., the von Neumann entropy of the reduced density matrix (\ref{rdmg}).

Let $\Omega$ and $\Lambda \subset \Omega$ be the spatial domains occupied by $\mathfrak{S}$  and  $\mathfrak{B}$  and $N$ and $L$ be the parameters determining the size of   $\mathfrak{S}$  and  $\mathfrak{B}$ (e.g. the corresponding side lengths if  $\Omega$ is a cube in $\mathbb{R}^d$ and $\Lambda$ is a sub-cube, so that $|\Omega|=N^d$ and $|\Lambda|=L^d$). We will assume that%
\begin{equation}
1\ll L \lesssim N,  \label{bip}
\end{equation}%
i.e., that $\mathfrak{E}$ (an "environment") is much larger
than $\mathfrak{B}$ (block) which can also be sufficiently large. 

The goal is to find the asymptotic form of $S_{\mathfrak{B}}$ in a certain
formalization of the heuristic inequalities (\ref{bip}).


Then the most widely used formalization of (\ref{bip}) is as follows. The r.h.s. of (\ref{bip}) is implemented in its strong form $L\ll$ N  via the
macroscopic limit $N\rightarrow \infty $ for $\mathfrak{S}$ in (\ref{SBE})
keeping $L$ fixed under a condition guarantying the existence of a well
defined limiting entanglement entropy
\begin{equation}
S_{\mathfrak{B}}=\lim_{N \to \infty}S_{\mathfrak{B} \mathfrak{S}}.
\label{sla}
\end{equation}%
Then the l.h.s. $1\ll L$ of (\ref{bip}) is implemented as the asymptotic
regime $L\rightarrow \infty $ for $S_{\mathfrak{B} }$, i.e., shortly
\begin{equation}
\text{first}\;N\rightarrow \infty, \; \text{then}\;L\rightarrow \infty .
\label{ar1}
\end{equation}

This asymptotic regime of the \emph{successive} limits has been considered in the
large number of works dealing with a variety of models of quantum gravity,
quantum field theory, quantum statistical mechanics and quantum information
science, see e.g.
\cite{Ab-Co:19,Al-Co:21,Am-Co:08,Ca-Co:11,Ho-Co:09,Pa:05,Wi:18,Ei-Co:11,La:15,Pe-Ei:09,Be-Co:20} for reviews.
It was found on the various levels of rigor that in the case of translation
invariant systems with short-range interaction and/or hopping the leading term of the large-$L$ asymptotic form of the macroscopic limit (\ref{sla}) of the entanglement
entropy (\ref{sla}) can be:

\smallskip
(i) the area law%
\begin{equation}
S_{\mathfrak{B} }=C_{d}^{\prime }\ L^{d-1}(1+o(1)),\;L\rightarrow \infty ,
\label{area}
\end{equation}%
if  $\mathfrak{S}$  is in its ground state which is not
critical (no quantum phase transition) or/and if there is a spectral gap
between the ground state and the rest of the spectrum;

\smallskip (ii) the enhanced (violation of) area law%
\begin{equation}
S_{\mathfrak{B} }=C_{d}^{\prime \prime }\ L^{d-1}\log L(1+o(1)),\;L\rightarrow
\infty ,  \label{viol}
\end{equation}%
if  $\mathfrak{S}$  is in its ground state which is critical
(a quantum phase transition is present);

\smallskip (iii) the volume law%
\begin{equation}
S_{\mathfrak{B} }=C_{d}^{\prime \prime \prime }\ L^{d}(1+o(1)),\;L\rightarrow
\infty,  
\label{vol}
\end{equation}%

%
%
%
if  $\mathfrak{S}$  is either in a mixed state, say, the Gibbs state of non-zero
temperature, or in a pure but sufficiently highly excited state, the latter
case is closely related to the fundamental Entanglement Thermalization
Hypothesis \cite{Ab-Co:19,De:18}. Note that the coefficients $C'_d, C''_d$, and $C'''_d$ do not depend on $L$.

Certain disordered quantum systems have also been considered, mainly various
spin chains, and both the one-dimensional area law and the enhanced area law
have been found and analyzed, see e.g. \cite{Ho-Co:09,Re-Mo:09,Ei-Co:11,St-Ab:14,Mi-Co:17} and references therein.

An argument establishing the above results turned out to be rather involved
and not always sufficiently transparent and undoubted, especially in the
multidimensional case. This is why a rather simple but non-trivial model of
free fermions living on the lattice $\mathbb{Z}^{d}$ has attracted a
considerable attention, see e.g. \cite{Pe-Ei:09,El-Co:16,Le-Co:17,Ly-Co:20,Pa-Sl:18} and references therein.

The model is described by the quadratic many-body Hamiltonian
\begin{equation}
\sum_{m,n\in \Omega }H_{mn}c_{m}^{+}c_{n},  \label{mbh}
\end{equation}%
where $\{c_{m},c_{m}^{+}\}_{m\in \Omega
},\;c_{m}^{+}c_{n}+c_{n}c_{m}^{+}=\delta _{mn}$ are the annihilation and
creation operators of free spinless fermions and
\begin{equation}
H_{\mathfrak{S} }=\{H_{mn}\}_{m,n\in \Omega }  \label{obh}
\end{equation}%
is their one-body Hamiltonian. Note that $H_\mathfrak{S}$ acts in the $|\Omega
|=N^{d}$ dimensional complex Euclidean space $\mathbb{C}^{|\Omega |}$, while
(\ref{mbh}) acts in the much "bigger" space $\mathcal{H}_{\mathfrak{S} }$ of
dimension $2^{|\Omega |}$, see (\ref{stsp}). The entries $\{H_{mn}\}$ of $H_\mathfrak{S}$ are sometimes called
hopping parameters.

It should be noted that the bipartite setting based on the form (%
\ref{stsp}) of the state space, which is widely used in quantum information
science (dealing with qubits) and quantum statistical physics (dealing with
spins), is not directly applicable to indistinguishable particles, fermions
in particular. Therefore, in this case, one proceeds not from states (see (%
\ref{stsp})), but from the algebra of observables of the entire system and
that (local) of its subsystems generated by the creation and annihilation
operators in the coordinate representation of the second quantization, see
e.g. \cite{Sz-Co:20,Wi:18,Be-Co:20} for reviews.

An important fact that facilitate strongly the analysis of the entanglement entropy of free fermions is   a convenient formula for $S_{\mathfrak{B}\mathfrak{S} }$ of (\ref{ente}%
)  expressing it via the so-called Fermi
projection of the one-body Hamiltonian (\ref{obh}), 
see e.g. \cite{Pe-Ei:09,St-Ab:14,Pe:03} and formulas (\ref{SBD}) -- (\ref{h}%
) below.
 The formula is as follows.

Given a point $\varepsilon_F$ on the spectral axis of $H_{\mathfrak{S} }$ denote $\chi _{ \varepsilon _{F}}$ the indicator
of $(- \infty,\varepsilon _{F}]$. Then
\begin{equation}
P_{\mathfrak{S} }=\chi _{\varepsilon _{F}}(H_{\mathfrak{S}
})  \label{PD}
\end{equation}%
is the Fermi projection of $H_{\mathfrak{S} }$ and $\varepsilon_F$ is the Fermi energy (a free parameter). It is the orthogonal projection on
the subspace of the one-body state space $\mathbb{C}^{|\Omega |}$ spanned by the eigenvectors $\{\psi
^{\alpha }\}_{\alpha=1}^{N}$ of $H_{\mathfrak{S} }$ with eigenvalues $%
\{\varepsilon _{\alpha }\}_{\alpha=1}^{N}$ belonging to $(
-\infty,\varepsilon _{F}]$, hence,
\begin{equation}
P_{\mathfrak{S} }=\{P_{m_{1}m_{2}}\}_{m_{1},m_{2}\in \Omega
},\;P_{m_{1}m_{2}}=\sum_{\varepsilon _{\alpha }\in \lbrack \varepsilon
_{0},\varepsilon _{F})}\psi _{m_{1}}^{\alpha }\;\overline{\psi
_{m_{2}}^{\alpha }}.  \label{fp}
\end{equation}%
Let
\begin{equation}
P_{\mathfrak{B}\mathfrak{S}  }=\{P_{l_{1}l_{2}}\}_{l_{1},l_{2}\in \Lambda }  \label{fpr}
\end{equation}%
be the restriction of $P_{\mathfrak{S} }$ to $\Lambda \subset \Omega $. Then we
have the formula \cite{Pe-Ei:09,Pe:03}: \begin{equation}
S_{\mathfrak{B} \mathfrak{S}}=\Tr_{L }h(P_{\mathfrak{B} \mathfrak{S} }),  \label{SBD}
\end{equation}%
where
\begin{equation}
h(x)=-x\log_2 x-(1-x)\log_2 (1-x),\;x\in \lbrack 0,1],  \label{h}
\end{equation}%
(the binary Shannon entropy) and $\Tr_{\Lambda }$ is the "partial" trace in $%
\mathbb{C}^{|\Lambda |}\subset \mathbb{C}^{|\Omega |}$ (do not mix it with $\mathrm{tr}_{\mathfrak{B}}$
in (\ref{rdmg}), (\ref{ente}), the trace operation in the $%
2^{|\Lambda |}$-dimensional state space $\mathcal{H}_{%
\mathfrak{B}}$ of the block in (\ref{stsp})).

The formula (\ref{SBD})  reduces the analysis of the
entanglement entropy of free fermions to the spectral analysis of the
one-body Hamiltonian $H_{\mathfrak{S} }$ of (\ref{mbh}) -- (\ref{obh}).

One more interesting aspect of the formula is that it provides a link with
the studies of asymptotic trace formulas for various classes of
matrix and integral operators, in particular, the so-called Szego's theorem and its
generalizations, see e.g. \cite{Le-Co:17,Pa-Sh:18, Ki-Pa:15, Bo-Si:90}
and references therein.

It is usually assumed that there exists a well defined infinite volume
Hamiltonian $H $ (cf. (\ref{sla}))%
\begin{equation}
H:=\lim_{N \to \infty}H_{\mathfrak{S} }  \label{H}
\end{equation}%
in a certain sense. In fact, this assumption is a weak form of the requirement for the
one-body Hamiltonian to have the short-range hopping and is quite natural in the regime (%
\ref{ar1}).

It follows then from the variety of works that in the translation invariant
case with a short-range hopping  $H$ (e.g. the discrete Laplacian) the leading term of the
asymptotic formula for the entanglement entropy in the regime (\ref{ar1})
have again one of the three forms (\ref{area}) -- (\ref{vol}).

Namely, it
is the area law (\ref{area}) if the Fermi energy $\varepsilon _{F}$ is in a
gap of the spectrum of $H$ of (\ref{H}), the enhanced area law (\ref{viol}%
) if $\varepsilon _{F}$ is in the spectrum of $H$,  and the
system in its ground state, i.e., at zero temperature. If, however, $T>0$,
hence, the indicator $\chi _{\lbrack \varepsilon _{0},\varepsilon _{F}]}$ in
(\ref{PD}) is replaced by the Fermi distribution
\begin{equation}
\left(1+e^{(\varepsilon -\varepsilon _{F})/T}\right)^{-1},\;T>0,  \label{fedi}
\end{equation}%
or, more generally, just by a continuous function, then we have the volume
law (\ref{vol}), see e.g. \cite{Am-Co:08,Ei-Co:11,La:15,Bi-Co:22} for
reviews.

For disordered free fermions, where the one-body Hamiltonian is the discrete
Schrodin\-ger operator with random potential (Anderson model), i.e., for the disordered short-range hopping case, the validity of
all three asymptotic formulas (\ref{area}) -- (\ref{vol}) for
the entanglement entropy has been rigorously established in \cite%
{El-Co:16,Pa-Sl:18,Pa-Sh:18,Mu-Co:20}. However, in this case the
area law is valid not only if the Fermi energy $\varepsilon _{F}$ is in the
gap of the spectrum of $H$ in (\ref{H}), but also if $\varepsilon _{F}$ in
the localized part of the spectrum. As for the validity of the enhanced area
law, it is the case if the Fermi energy coincides with a so-called
transparency energy of $H$, see \cite{ Mu-Co:20} for  this result
 and \cite{LGP}, Section 10.3 for the definition and
properties of transparency energies.

In addition, certain new properties of the entanglement entropy were found
in the disordered case: the vanishing of the fluctuations of the
entanglement entropy (selfaveraging) for $d\geq 2$ as $L \to \infty $ \cite%
{El-Co:16}, nontrivial fluctuations for $d=1$ \cite{Pa-Sl:18}, the Central
Limit Theorem for the entanglement entropy at nonzero temperature for $d=1$
(see (\ref{fedi})) and, as a result, the $L^{1/2}$ (instead of $L^{0}$)
scaling of the sub-leading term for the volume law for $d=1$ \cite{Pa-Sl:18}.

As already mentioned, the above asymptotic results for both spin systems and free fermions were
obtained in the  successive limits regime (\ref{ar1}). On the other hand, one
can consider the implementations of heuristic inequalities (\ref{bip}) where
$N$ and $L$ tend to infinity \emph{ simultaneously} 
\begin{equation}
N\rightarrow \infty ,\;L\rightarrow \infty ,\;L/N^{\alpha }\rightarrow
\lambda _{\alpha }>0,\;\alpha \in (0,1],\; \lambda_{1} \in (0,1],
\label{diff}
\end{equation}%
e.g.,
\begin{equation}
L=[\lambda _{\alpha }N^{\alpha }]=\lambda _{\alpha }N^{\alpha
}+O(1),\;N\rightarrow \infty.  \label{diffa}
\end{equation}

The both asymptotic regimes (\ref{ar1}) and (\ref{diff}) -- (\ref{diffa})
are of interest in view of the general bipartite setting (\ref{SBE}) -- (\ref%
{bip}). In addition,  the double scaling regimes (\ref{diff}) -- (\ref{diffa}%
) are  important because they  seem more adequate to a wide variety of
numerical studies of entanglement in extended systems, where it is often hard, if not possible, to
implement appropriately the successive limits (\ref{ar1}).

The regime (\ref{diff}) -- (\ref{diffa}%
) for a short-range hopping case is considered in \cite{Pa-Sl:24}
where it is shown that if $H$ is the one-dimensional discrete Laplacian, then the enhanced area law ((\ref{viol}) with $d=1$) is valid if $0 < \alpha < 2/3$, thereby manifesting a certain universality of this asymptotic form with respect to the scaling of the block.

\medskip In this paper we consider  the regime (\ref{diff}) -- (\ref{diffa}) for
the systems of free fermions where $H_{\mathfrak{S} }$ is the $N\times N$
hermitian random matrix having a unitary invariant probability law, e.g. the well known Gaussian Unitary Ensemble (GUE), see \cite%
{Pa-Sh:11} for results and references. It is widely believed and confirmed
by various recent results (see, e.g. \cite{Pa-Sh:11,Er:11}) that large
random matrices may model multi-component and multi-connected media playing the role of the mean field type approximation for
the Schrodinger operator with random potential, a basic model in the theory
of disordered systems and related branches of spectral theory and solid
state theory.
We will show that in the case of
this (long-range) one-body Hamiltonian the entanglement entropy obey the
volume law (\ref{vol}) with $d=1$, see Result \ref{r:rmas}.

The corresponding results are presented in Section 2.1 and 2.2 and are
proved in the Appendices \ref{a:hbou} -- \ref{a:L1}.

In Section \ref{ss:page} we  deal with a related problem although it does not involve free fermions. The problem was initially
considered in the context of quantum gravity, where the roles of $\mathfrak{E}$
and $\mathfrak{B}$ in (\ref{SBE}) play a black hole in the  pure initial
state of the evaporation process and outgoing Hawking radiation
respectively \cite{Pa:05,Pa:93a,Pa:93b}.
The idea was that the generic
evaporative dynamics of a black hole may be captured by the random sampling
of subsystems of a  quantum system which is in a pure random initial state.

This was one of the first applications of random matrices to cosmology that
prompted extensive activities covering several fields, see e.g. \cite%
{Al-Co:21,Da-Co:14,Pa:05,Bi-Co:22} for reviews.

It is also worth mentioning that there is a link of the results of \cite{Pa:05,Pa:93a,Pa:93b} with the asymptotic
formulas (\ref{area}) -- (\ref{vol}), especially with the volume law (see Section \ref{ss:page}).

\section{Results}
\label{s:res}
We present here our results and their discussions. The corresponding technical proofs are given in Appendices \ref{a:hbou} -- \ref{a:page}.

\subsection{Generalities}
\label{ss:gen}

To study possible asymptotic formulas for the entanglement entropy (\ref{SBD}%
) of free fermions at zero temperature, we will use general bounds given by%
%
%

\begin{result}
\label{r:1} Given the general setting (\ref{mbh}) -- (\ref{h}) for the model
of free lattice fermions, we have
the following bounds for the entanglement entropy $S_{\mathfrak{S}\mathfrak{B}}$ (\ref{ente}):
\begin{equation}
\mathcal{L}_{\mathfrak{S}\mathfrak{B} }\leq S_{\mathfrak{S}\mathfrak{B} }\leq \mathcal{U}%
_{\mathfrak{S}\mathfrak{B} },  \label{tsb}
\end{equation}%
where%
\begin{align}\label{lub}
&\mathcal{L}_{\mathfrak{S}\mathfrak{B} }=4\Tr_{\Lambda} P_{\mathfrak{S}\mathfrak{B} }(\mathbf{1}%
_{\Lambda }-P_{\mathfrak{S}\mathfrak{B} })=\sum_{l \in \Lambda, k \in \Omega \setminus \Lambda} |P_{lk}|^2,
\\
&\hspace{2cm}\mathcal{U}_{\mathfrak{S}\mathfrak{B}}=|\Lambda |
\ h_{0}(\mathcal{L}_{\mathfrak{S}\mathfrak{B} }/4|\Lambda |), \notag
\end{align}
and
\begin{equation}
h_{0}:[0,1/4]\rightarrow \lbrack 0,1],\;h(x)=h_{0}(x(1-x)),\;x\in \lbrack
0,1]
\label{h0h}
\end{equation}%
%
with $H$ given by (\ref{h}).

If the one-body Hamiltonian $H_\mathfrak{S}$ is random, then (\ref{tsb}) -- (\ref{lub}) are valid for every realization, while  we have for the expectation $\mathbf{E}\{S_{\mathfrak{S}\mathfrak{B} }\}$
of $S_{\mathfrak{S}\mathfrak{B} }$
\begin{equation}
\overline{\mathcal{L}}_{\mathfrak{S}\mathfrak{B} }\leq \mathbf{E}\{S_{\mathfrak{S}\mathfrak{B} }\}\leq \overline{\mathcal{U}}%
_{\mathfrak{S}\mathfrak{B} },  \label{tsbr}
\end{equation}
where
\begin{equation}
\overline{\mathcal{L}}_{\mathfrak{S}\mathfrak{B} } =\mathbf{E}\{\mathcal{L}_{\mathfrak{S}\mathfrak{B} }\}, \; \overline{\mathcal{U}}
_{\mathfrak{S}\mathfrak{B} }=|\Lambda |h_0(\overline{\mathcal{L}}_{\mathfrak{S}\mathfrak{B} }/4|\Lambda |).  \label{lubr}
\end{equation}

\end{result}

The bounds are proved in Appendix \ref{a:hbou}. They allow us to obtain,
by using  an elementary argument and technique, rather tight bounds for  the entanglement entropy,  see Figures \ref{fig1} -- \ref{fig3} and Table 1 below.   

The same bound are used in \cite{Pa-Sl:24} to study the enhances area law (\ref{viol}) for translation invariant free fermions in the regime (\ref{diff}) -- (\ref{diffa}).


Note also that the lower bound $\mathcal{L}_{\mathfrak{S}\mathfrak{B}}$ in (\ref{tsb}) -- (%
\ref{lub}) is proportional to the variance of the number of fermions in $%
\Lambda $, see e.g. \cite{Mi-Co:17} and references therein. This
quantity can also be expressed via the density-density correlator $(\delta
(E^{\prime }-H_{\mathfrak{S} }))_{mn}(\delta (E^{\prime \prime }-H_{\mathfrak{S}
}))_{nm} $, important in the solid state theory \cite{LGP,Ki-Co:03}.

For other
versions of two-sided bounds for the entanglement entropy see \cite%
{Ei-Co:11,Bi-Co:22,He-Co:11,Pa-Sl:14,Wo:08}) and references therein.

We will also use the spectral version of the basic formula (\ref{SBD}) -- (%
\ref{h}). Let
\begin{equation}
\mathcal{N}_{P_{\mathfrak{B}\mathfrak{S} }}=\sum_{\alpha=1}^{|\Lambda|}\delta _{p_{\alpha }}
\label{dosp}
\end{equation}%
be the counting measure of eigenvalues $\{p_{\alpha }\}_{\alpha=1}^{|\Lambda|}$ of
 $P_{{\mathfrak{B}\mathfrak{S} }}$ of (\ref{fpr}). Then
we can write (\ref{SBD}) as%
\begin{equation}
S_{{\mathfrak{B}\mathfrak{S} }}=\int_{0}^{1}h(p)\mathcal{N}_{P_{{\mathfrak{B}\mathfrak{S} } }}(dp).
\label{slgen}
\end{equation}%
This reduces the asymptotic study of $S_{{\mathfrak{B}\mathfrak{S} } }$ to that of $%
\mathcal{N}_{P_{{\mathfrak{B}\mathfrak{S} } }}$. The latter is often not simple to find
(see, however, \cite{La-Wi:80},  and also (\ref%
{mp1}) -- (\ref{mop}), and   (\ref%
{wach}) and (\ref{rho}) below), but formula (\ref{slgen}) proves to be also useful to
interpret various results on the entanglement entropy of free fermions.


\subsection{Entanglement Entropy of Free Fermions with a Random Matrix Hamiltonian}
\label{ss:rm}

We will assume here that the whole system $\mathfrak{S}$
and its block ${\mathfrak{B} }$ occupy the integer valued intervals
\begin{equation}
\Omega =(1,2,\dots,N), \; \Lambda =(1,2,\dots,L).
\label{larmt}
\end{equation}
It is convenient at this point  to change the notation and write subindices $N$ and $L$ instead of $\mathfrak{S}$ and $\mathfrak{B}$:
\begin{equation}
\mathfrak{S} \to N, \; \mathfrak{B} \to L.  \label{frlat}
\end{equation}
We will assume then that
the one-body
Hamiltonian (\ref{obh}) is
\begin{equation}
H_{N }=M_{N},  \label{hamrmt}
\end{equation}%
where $M_{N}$ is the $N\times N$ hermitian random matrix whose probability
law is invariant with respect to the all unitary transformations $%
M_{N}\rightarrow U_{N}M_{N}U_{N}^{\ast }, \; U_N \in U(N)$. An interesting and widely studied subclass  of this class
of random matrices  consists of the so-called matrix models (also known as invariant ensembles), where the matrix
probability law is%
\begin{align}
& Z_{N}^{-1}e^{-N\mathrm{Tr}V(M_{N})}dM_{N},  \notag \\
V(x)& \geq (1+\varepsilon )\log (1+x^{2}),\;\varepsilon >0,\;x\in \mathbb{R,} \notag \\
& \hspace{-1.5cm}dM_{N}=\prod_{1\leq n\leq N}dM_{nn}\prod_{1\leq n_{1}<n_{2}\leq N}d\Re
M_{n_{1}n_{2}}d\Im M_{n_{1}n_{2}},
\label{GUE}
\end{align}%
see, e.g. \cite{Pa-Sh:11,Br-Wa:93}.

The most known example of matrix models is the Gaussian Unitary Ensemble
(GUE), where $V(x)=2x^{2}/\varepsilon_0$. In this case the entries of $M_{N}$
are complex Gaussian random variables:
\begin{align}
M_{N}&= \varepsilon _{0}(4N)^{-1/2}\{X_{m_{1}m_{2}}\}_{m_{1},m_{2}=1}^{N},
\notag \\
\mathbf{E}\{X_{m_{1}m_{2}}\}&= \mathbf{E}\{X_{m_{1}m_{2}}^{2}\}=0,\;\mathbf{E%
}\{|X_{m_{1}m_{2}}|^{2}\}=1.  \label{gue}
\end{align}%
Thus, the entries of $M_{N}$  in (\ref{GUE}) -- (\ref{gue}) have the same order of magnitude ($N^{-1/2}$ for the GUE), hence, the limit operator $H$ of (\ref{H}) does not exist in this case. This should be contrasted with the  short-range hopping case where the limiting operator is well defined and is a discrete Laplacian in the simplest case of lattice translation invariant fermions, and a Schrodinger operator with random potential (Anderson model) for disordered free fermions both acting in $l^2(\mathbb{Z}^d)$, see e.g. \cite{El-Co:16} and references therein.

An analogous situation is in the mean field models of statistical mechanics. On the other hand, a number of important characteristics have well defined macroscopic
limits (e.g. the free energy in statistical mechanics, and the limiting Normalized
Counting Measure
in
random matrix theory). This allow us to view (\ref{hamrmt}) as
a disordered version of the mean field model for free fermions and to expect
that the entanglement entropy have a well defined asymptotic behavior in
this case.

Note that $M_{N}$ of (\ref{gue}), more precisely, its real symmetric analog (GOE),
is used as the interaction matrix in a highly non-trivial mean field model
of spin glasses known as the Scherrington-Kirkpatrick model \cite{Me-Co:86},
where the role of Fermi operators in (\ref{mbh}) play classical or quantum
spins. The model is a disordered version of the well known Kac model where
the interaction matrix is $\varepsilon_{0} N^{-1}\mathbf{1}_{N}, \; \varepsilon_{0} >0$ and
the corresponding spin model reproduces the well known Curie-Weiss description of the ferromagnetic
phase transition in the large-$N$ limit.

By the way, by using the "Kac" interaction matrix $\varepsilon_{0}
N^{-1}\{1\}_{m_{1},m_{2}=1}, \; \varepsilon_{0} >0$ as the one-body Hamiltonian
in (\ref{mbh}), it is easy to find that the corresponding entanglement
entropy is independent of $L$. Indeed, write
\begin{equation}  \label{kac}
H_{N}=\varepsilon_{0}  N^{-1}\mathbf{1}_{N}, \;  =\varepsilon_{0}
P_{d_{\Omega }},
\end{equation}
where $P_{d_{\Omega }}$ is the orthogonal projection on the "diagonal" vector $%
d_{\Omega }=|\Omega |^{-1/2}\{1,\ldots ,1\}\in \mathbb{C}^{|\Omega |}$.

Then
the corresponding Fermi projection is (see (\ref{PD}) -- (\ref{fp}) and (\ref{frlat}))
\begin{equation*}
P_{ N }=\chi _{(-\infty ,\;\varepsilon _{F}]}(0)(\mathbf{1}_{\Omega
}-P_{d_{\Omega }})+\chi _{(-\infty ,\;\varepsilon _{F}]}(\varepsilon_{0}
)P_{d_{\Omega }},
\end{equation*}
where $\chi _{(-\infty, \; \varepsilon _{F}]}$ is the indicator of $(-\infty
,\varepsilon _{F}]\subset \mathbb{R}$, and the restriction (\ref{fpr}) of $P_{ \Omega }$ to $\Lambda=[1,\dots,L]$ is
\begin{equation*}
P_{LN }=\chi _{(-\infty ,\;\varepsilon _{F})}(0)(\mathbf{1}%
_{\Lambda }-L/N P_{d_{\Lambda }})+\chi _{(-\infty ,\;\varepsilon
_{F})}(\varepsilon_{0})L/N P_{d_{\Omega }}\chi _{(-\infty ,\;\varepsilon
_{F})}(\varepsilon_0 ).
\end{equation*}%
%
%
It follows then from (\ref{SBD}) that%
\begin{equation}
\label{eekac}
S_{LN }=h(L/N)=h(1-L/N).
\end{equation}
Hence, the entanglement entropy is
zero in the regime (\ref{ar1}) of successive limits (moreover, $S_{\Lambda }$
of (\ref{sla}) is already zero), and in the regime (\ref{diff}) of
simultaneous limits if $\alpha <1$, while it is%
\begin{equation}
\lim_{N \to \infty, L \to \infty, L/N \to \lambda_1} S_{LN }=h(1-\lambda_{1} )=h(\lambda _{1})  
\label{kacee}
\end{equation}%
in the regime (\ref{diff}) with $\alpha =1$, i.e., for an asymptotically
proportional $|\Lambda |=L$ and $|\Omega |=N$.

Formula (\ref{kacee}) corresponds formally to the one-dimensional area law (%
\ref{area}), although the notion of surface is not well defined in the mean
field setting.

\smallskip

We will show now that for random matrices (\ref{GUE}), a
disordered version of the Kac model, the situation is in some sense
"opposite", since in this case the entanglement entropy obeys the analog of
the volume law (\ref{vol}).

To this end we note first that because of the unitary invariance of (\ref%
{GUE}) the eigenvalues and the eigenvectors of $M_{N}$ are
statistically independent and eigenvectors form a random unitary matrix $%
U_{N}=\{U_{jk}\}_{j,k=1}^{N}$ that is uniformly (Haar) distributed over the
group $U(N)$ \cite{Pa-Sh:11}. 
Hence, the Fermi projection (\ref{PD}) -- (\ref{fp}) in this case is (see (\ref{frlat})
\begin{equation}
(P_{N})_{m_{1}m_{2}}=\sum_{k=1}^{K}U_{m_{1}k}\overline{U_{m_{2}k}},\,m_{1},m_{2}=1,\ldots ,N.  \label{pdma}
\end{equation}%
where%
\begin{equation}
K=[N\kappa _{F}],\;\kappa_{F}=\nu _{M}(\varepsilon_{F}) \in (0,1),
\label{Mma}
\end{equation}%
$\kappa _{F}$ is the analog of the Fermi momentum fixing the ground state (the Fermi sea) of free fermions, and $\nu _{M}$ is the
limiting Normalized Counting Measure of $M_{N}$ (cf. (\ref{dosp}))
\begin{equation}
\nu _{M}=\lim_{N\rightarrow \infty }N^{-1}\mathcal{N}_{M_{N}},  \label{numl}
\end{equation}%
see \cite{Pa-Sh:11,Br-Wa:93} for the proof of (\ref{numl}) and various
examples, the most known is the Wigner semicircle law
\begin{equation*}
\nu^{\prime
}_M(\varepsilon)=2(\pi \varepsilon_0^2 )^{-1} (\varepsilon_0^2
-\varepsilon^2)^{1/2}
\chi_{[-\varepsilon_0,\varepsilon_0]}(\varepsilon)
\end{equation*}
for
the GUE (\ref{gue}). 

Furthermore, the analog of the restriction $P_{LN }$ (\ref{fpr})
of $P_{N }$ (\ref{pdma}) is in this case%
\begin{equation}
(P_{LN})_{l_{1}l_{2}}=\sum_{k=1}^{K}U_{l_{1}k}\overline{U_{l_{2}k}},\,l_{1},l_{2}=1,\ldots ,L.  \label{pdlma}
\end{equation}%
We will again begin with the asymptotic bounds (\ref{tsb}), this time for
the expectation $\mathbf{E}\{S_{LN }\}$ of the entanglement
entropy (\ref{SBD}) corresponding to (\ref{hamrmt}).

To simplify the further notation we will write below $\kappa$ instead of $\kappa_F $ (see (\ref{Mma})), and $ \lambda$
instead of $\lambda_1$ (see (\ref{diff}) -- (\ref{diffa}), (\ref{kacee})):
\begin{equation}\label{laka}
\lambda_1 \to \lambda, \; \kappa_F \to \kappa.
\end{equation}

\begin{result}
\label{r:rmlob} Let the one-body Hamiltonian $H_{N }$ of the system of
free fermions be the random matrix (\ref{hamrmt}). Assume that
(see (\ref{Mma}))
\begin{align}
& \hspace{1cm}N\rightarrow \infty ,\;K\rightarrow \infty ,\;L\rightarrow
\infty, \; K/N\rightarrow \kappa\in (0,1).  \label{kln}
\end{align}%
Then the expectation of the entanglement entropy (see (\ref{SBD}) -- (\ref{h}%
)) of the block (\ref{larmt})\ admits the asymptotic bounds
\begin{align}
&\hspace{1.5cm} \widehat{\mathcal{L}}_{LN}\leq \mathbf{E}%
\{S_{LN}\}\leq \widehat{\mathcal{U}}_{LN},  \notag \\
& \widehat{\mathcal{L}}_{LN}=4\kappa(1-\kappa)L(1-L/N)+
o(L),  \notag \\
&\hspace{-0.5cm}\widehat{\mathcal{U}}_{LN}=L h_0 (\kappa(1-\kappa)L(1-L/N)) + o(L).  \label{lbrmt}
\end{align}
\end{result}
The proof of the result is given in Appendix \ref{a:m4}.

\smallskip The final asymptotic bounds for $\mathbf{E}\{S_{LN}\}$
are determined by the order of magnitude of $L$ with respect to $N$ as $N\rightarrow
\infty$, see (\ref{diff}).

\smallskip (i) $N \to \infty, \; L \to \infty, \; L/N \to 0$, i.e., $0 < \alpha <1 $ in (\ref%
{diff}) -- (\ref{diffa}):
\begin{align}
&C_{-} L+o(L)\leq \mathbf{E}\{S_{LN}\}\leq C_{+} L+ o(L),  \notag \\
&C_{-}=4\kappa (1-\kappa ), \; C_{+}=h_{0}(\kappa(1-\kappa)).  
\label{am1}
\end{align}
Moreover, since $C_-=C_+=1$ for $\kappa =1/2 $ (see (\ref{h}) and (\ref{h0h})), we
have in this case an exact asymptotic formula
\begin{equation}  \label{ee12}
\mathbf{E}\{S_{LN}\}= L+ o(L), \;L\rightarrow \infty.
\end{equation}
Note that these bounds are valid even for a finite $L$, but with the
replacements of $o(L)$ by $o(N)$.

Bearing in mind that $L$ plays in this case the role of the size (volume) of
the block, this result can be viewed as an indication of the validity of
the volume law (\ref{vol}) for the mean entanglement entropy, both in the
regime (\ref{ar1}) of simultaneous limits and in the regime (\ref{diff}) -- (%
\ref{diffa}) of successive limits for $\alpha <1$. Note that in the translation invariant short-range case we have in this situation
the enhanced area law (\ref{viol}), see \cite{Le-Co:17} and references therein.

\medskip (ii) $N \to \infty, \; L/N \to \lambda \in (0,1)$, i.e., $\alpha =1$ in (%
\ref{diff}) -- (\ref{diffa}), see (\ref{laka}):
\begin{align}
&C_{-} L+o(L)\leq \mathbf{E}\{S_{LN}\}\leq C_{+} L+ o(L),
\;L\rightarrow \infty  \notag \\
&C_{-}=4\kappa (1-\kappa )(1-\lambda), \; C_{+}=h_{0}(C_-/4).  \label{ae1}
\end{align}
The bounds rules out the area law (\ref{area}) and the enhanced
area law (\ref{viol}) and are compatible only with the volume law but with coefficient
different  from those of the previous case (\ref{am1}). In particular, since  $0 < \lambda \le 1$ (see (\ref{laka})) in general, an exact asymptotic formula
(see (\ref{ee12})) cannot be obtained from (\ref{ae1}) in general.

For similar results pertinent to related matrix models and their applications see \cite{Bi-Co:22}.

\smallskip We conclude that for the random matrix (long range) one-body
Hamiltonian (\ref{hamrmt}) of free fermions a possible
asymptotic law for the expectation of the entanglement entropy in the both
asymptotic regimes (\ref{ar1}) and (\ref{diff}) -- (\ref{diffa}) is an
analog of the volume law (\ref{vol}). 
\smallskip

To see the indications for other possible scalings of $\mathbf{E}%
\{S_{LN}\}$ let us consider the case where%
\begin{equation}
L=N(1-\delta _{N}),\;\delta _{N}=o(1),\;N\rightarrow \infty ,  \label{ldua}
\end{equation}
corresponding to the blocks with size $L$ close to the size $N$ of the
entire system. Then (\ref{lbrmt}) implies for, say, $\delta _{N}=\log N/N$%
\begin{equation}
4\kappa(1-\kappa)\log L(1+o(1))\le \mathbf{E}\{S_{LN}\}
\le \kappa(1-\kappa )\log ^{2}L(1+o(1)),\;N\rightarrow \infty ,
\label{sdua}
\end{equation}%
and we obtain the bounds that are compatible only with the one dimensional
enhanced area law scaling (\ref{viol}).

For similar bounds in the translation invariant short-range hopping case see \cite{Pa-Sl:24,Wo:08}.


\medskip We will now use certain random matrix theory results to show that in
the above case (ii)
of the asymptotic regime (\ref{diff}) -- (\ref{diffa}) of simultaneous
limits the analog of the volume law is valid for all typical realizations of
the entanglement entropy itself as well as for its expectation.

\begin{result}
\label{r:rmas} Under the conditions of previous Result \ref{r:rmlob}, i.e.,
for
\begin{equation}\label{KLN}
K \to \infty, \; L \to \infty, \; N \to \infty , \; K/N \to \kappa\in (0,1), \; L/N \to \lambda \in (0,1)
\end{equation}
the entanglement entropy (\ref{SBD}) -- (\ref{h}) of the block $\Lambda$ of (%
\ref{larmt}) admits the volume law asymptotic formula valid with probability
1
\begin{equation}
S_{LN}=Ls_{\kappa \lambda }+o(L),\;N\rightarrow \infty .
\label{sl1}
\end{equation}\label{sl2}%
Here the coefficient (the "specific" entropy) $s_{\kappa \lambda }$ is non-random and equals
\begin{equation}\label{specs}
s_{\kappa \lambda }=\int_{p_{-}}^{p_{+}}h(p)\nu _{ac }^{\prime }(p)dp
\end{equation}%
with
\begin{align}
& \hspace{1cm}\nu _{ac }^{\prime }(p)=\frac{\sqrt{(p_{+}-p)(p-p_{-})}}{2\pi \lambda p(1-p)}\chi_{[p_{+},p_{-}]}(p),\quad p_{\pm }=\left( \sqrt{\kappa
(1-\lambda )}\pm  \sqrt{\lambda(1-\kappa )}\right) ^{2}.  \label{rho}
\end{align}
\end{result}
\noindent The proof of the result is given in Appendix \ref{a:rmha}.

\smallskip

It is curious that if $\kappa=\lambda=1/2$, then $\nu^{\prime
}_{ac}(p)=\bigl(\pi \sqrt{p(1-p)}\bigr)^{-1/2}\chi_{[0,1]}(p)$,  the density of
the limiting Normalized Counting measure (the Density of States) of the
one-dimensional lattice Laplacian.

\begin{figure}[ht]
\center{\includegraphics[width=12cm]{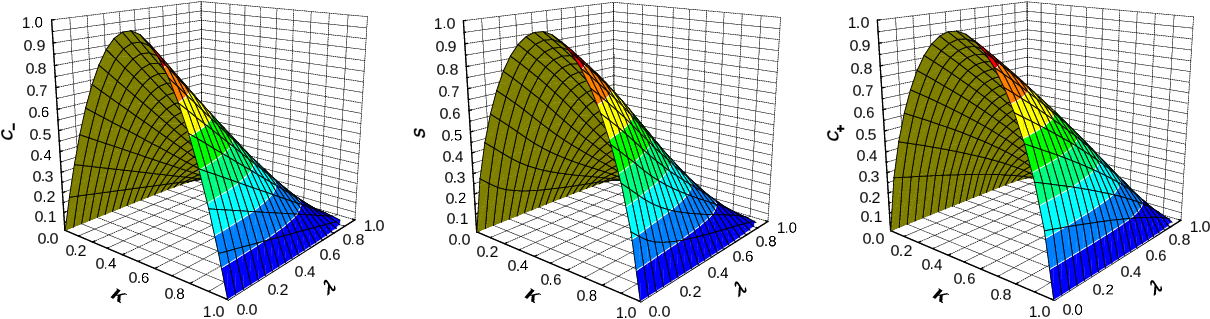}}
\caption{{\small The coefficients $C_-$ of (\ref{ae1}) (left panel), $s_{\kappa\lambda}$ of (\ref{specs}) -- (\ref{rho}) (central panel), and $C_+$ of (\ref{ae1}) (right panel)  as functions of parameters $\kappa$ of (\ref{Mma}) and $\lambda$ of (\ref{diffa}).}}
\label{fig1}
\end{figure}

\medskip
The coefficient $s_{\kappa\lambda}$ in (\ref{sl2})
is a complex function of $(\kappa,\lambda) \in [0,1]^2$, having different
expressions in four sectors of the square $[0,1]^2$. They are determined by the conditions on the
vanishing of atoms of the limiting Normalized Counting Measure $\nu_{\kappa,\lambda}$ of (\ref{ncmp}) -- (\ref{wach})
: $m_0=m_1=0$; $m_0=0, m>0$; $m_0>0, m_1=0;$
and $m_0,m_1>0$ (cf. (\ref{mul})). This is because the integral in (\ref{specs}) is equal to a rather involved combination of $\text{ext}(\lambda, \kappa)$ and $\text{ext}(\lambda, 1-\kappa)$ and their logarithms, where "$\text{ext}$" denotes either $\min$ or $\max$. The corresponding calculations and the result are similar to but more involved than those in Appendix \ref{a:page}, dealing with a one-parametric analog of the above. In particular, the plot on Figure \ref{fig4}  of the piece-wise analytic function, given by the second term in the r.h.s. in (\ref{mul}) (see also (\ref{mulrr})) is the one-parametric analog of Figure \ref{fig1} describing the surface 
$s_{\kappa\lambda}, \ (\kappa,\lambda) \in [0,1]^2$.

\begin{figure}[ht]
\center{\includegraphics[width=10cm]{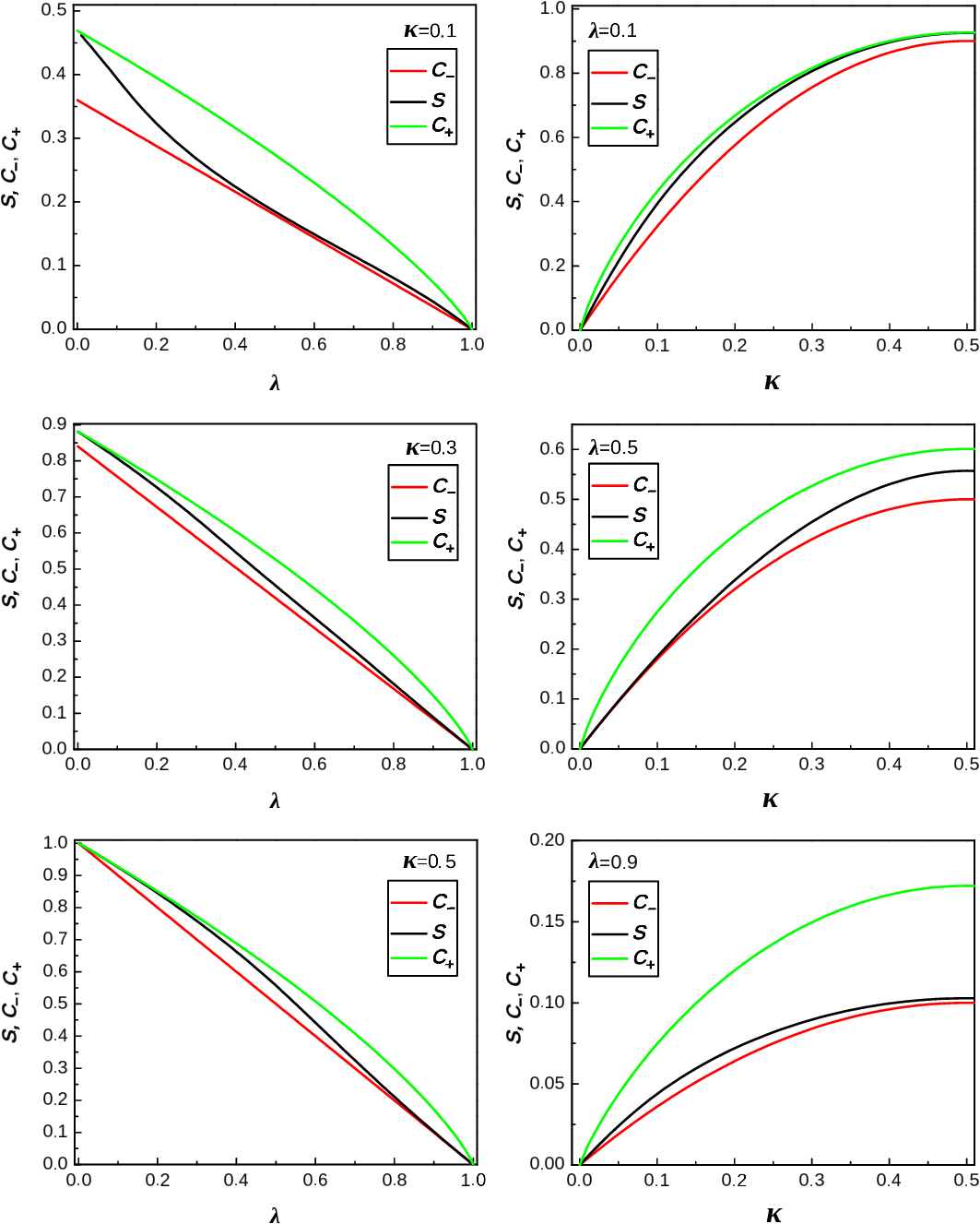}}
\caption{{\small Left: The coefficients $C_-$ (red) and $C_+$ (green) of (\ref{ae1}), and $s_{\kappa\lambda}$ (black) of (\ref{specs}) as functions $\lambda$ for different values of $\kappa$.
Right: The same coefficients as functions of $\kappa$ for different values of $\lambda$.}}
\label{fig2}
\end{figure}

\begin{figure}[ht]
\center{\includegraphics[width=12cm]{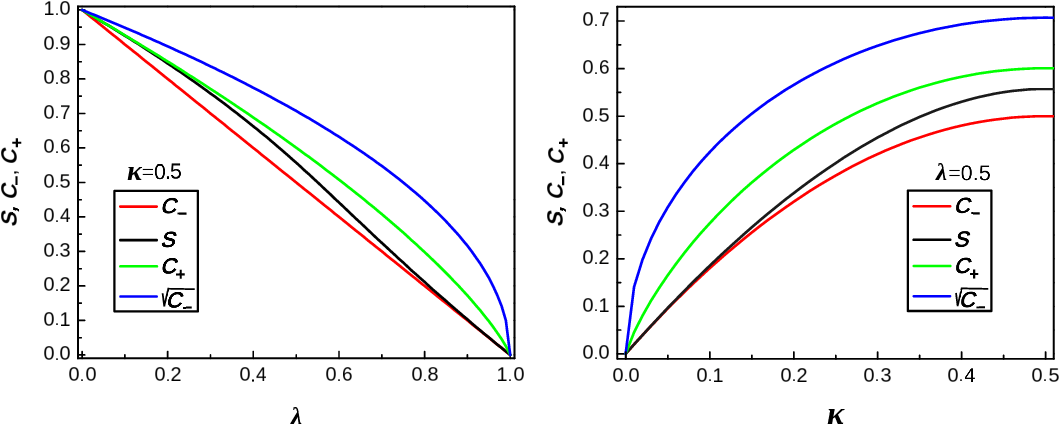}}
\caption{{\small The same as on Fig.~\ref{fig2} and the coefficient   $\sqrt{C_-}$ (blue) of (\ref{u1/2}).}}
\label{fig3}
\end{figure}

\medskip
This is why we will give below certain graphic and numeric results concerning the coefficient $s_{\kappa\lambda}$ in (\ref{sl1}) -- (\ref{rho}) and the coefficients $C_{\pm}$ in bounds (\ref{ae1}). Figures \ref{fig1} -- \ref{fig3}  present various graphic manifestation of proximity of $C_{-}$ and $C_{+}$ to $s_{\kappa \lambda}$ for the various pairs $(\kappa,\lambda) \in [0,1]^2$ of the parameters $\kappa$ and $\lambda$ of the Hamiltonian (see (\ref{Mma}) and (\ref{rho})).
Figure \ref{fig1} gives the shape of three "surfaces" describing $C_-$, $C_+$ and $s_{\kappa \lambda}$, which are quite close to each other. Note that the surface of the central  panel is the two-parameter analog of piece-wise analytic curve of Figure \ref{fig4} describing  (\ref{mul}).
Figure \ref{fig2} gives the values of the $C_-$, $C_+$ and $s_{\kappa \lambda}$ as functions of one of the  parameters for certain fixed values of the other, and Figure \ref{fig3} gives the same values, supplemented by those of the coefficient $C_-^{1/2}$ of the upper bound
\begin{equation}\label{u1/2}
\mathcal{U}^*_{\Lambda, \Omega}=(L\mathcal{L}_{\Lambda, \Omega})^{1/2}=C_-^{1/2}L+o(L),
\end{equation}
see \cite{Ei-Co:11,He-Co:11,Pa-Sl:14} and references therein concerning the bound.

Table 1
shows numerical data on the closeness of the curves of Figure \ref{fig2} measured by the maximum distances between the corresponding pairs of curves.

We conclude that in the asymptotic regime (\ref{kln}), known in random
matrix theory as the global regime, we have with probability 1 (for all
typical realizations) an analog of the volume law that is quite well approximated by
bounds given in (\ref{tsb}).


To see a possibility of other than the volume law asymptotic forms of the
entanglement entropy in the random matrix case, let us assume that $\lambda =1-\delta $ with a sufficiently small (but $N$-independent) $\delta >0$ corresponding to the blocks of size close to
that of the whole system (cf. (\ref{ldua})). It follows then from (\ref{rho}):%
\begin{equation*}
p_{\pm }|_{\lambda =1-\delta }=(1-\kappa ) \pm 2^{ }\,\delta ^{1/2}%
\sqrt{\kappa(1-\kappa)}+O(\delta ),
\end{equation*}%
hence, the width of the support of $\nu_{ ac}^{\prime }$ is $O(\delta
^{1/2}) $ and we obtain in view of (\ref{sl1}) -- (\ref{sl2}) and (\ref{rho})%
\begin{equation}
s_{\kappa, 1-\delta }=h(\kappa )\,\delta +o(\delta),\;\delta =o(1).
\label{ente1}
\end{equation}%
The last formulas can be interpreted as an indication of possibility to
obtain the scaling $L=o(N)$, i.e., a "subvolume" laws asymptotic formulas in
the random matrix case. Here is another indication provided by the case $%
L=N-1$, i.e., (\ref{ldua}) with $\delta_N=N^{-1}$. In this case it is possible to find an exact asymptotic formula valid with
probability exceeding $1-\varepsilon $ for any $\varepsilon >0$, i.e., for
the overwhelming majority of realizations (see Appendix \ref{a:L1}):
\begin{equation}
S_{LN}|_{L=N-1}=h(\kappa)+o(1),\;N\rightarrow \infty ,  \label{eel1}
\end{equation}%
i.e., we have a formal analog of the one-dimensional area law.

In particular (cf. (\ref{sl1}) and (\ref{ente1}))%
\begin{equation*}
S_{LN}|_{L=N-1,\kappa =1/2}=1+o(1),\;N\rightarrow \infty.
\end{equation*}%
In addition, we have in this case 
$(4\kappa(1-\kappa)+O(1/N))|_{\kappa=1/2}=1+O(1/N),$ i.e., the lower
bound (\ref{sdua}) of the entanglement entropy
coincides with its value for the overwhelming majority of realizations.

\bigskip

\begin{table}
\begin{minipage}[ht]{0.49\linewidth}
\begin{tabular}{|c|c|c|c|}
\hline
$\kappa$        & $\Delta(C_-,s)$       &       $\Delta(s, C_+)$ & $\Delta(C_-, C_+)$\\
\hline
0.1     & 0.10  & 0.09  & 0.1\\
0.2     & 0.08  & 0.096 & 0.1\\
0.3     & 0.05  & 0.08  & 0.1\\
0.4     & 0.05  & 0.08  & 0.1\\
0.5     & 0.06  & 0.09  & 0.1\\
0.6     & 0.06  & 0.08  & 0.1\\
0.7     & 0.05  & 0.08  & 0.1\\
0.8     & 0.08  & 0.09  & 0.1\\
0.9     & 0.1   & 0.09  & 0.1\\
\hline
\end{tabular}
\label{Table1}
\end{minipage}
\hfill
\begin{minipage}[ht]{0.49\linewidth}
\begin{tabular}{|c|c|c|c|}
\hline
$\lambda$       & $\Delta(C_-,s)$       &       $\Delta(s, C_+)$ & $\Delta(C_-, C_+)$\\
\hline
0.1     & 0.07  & 0.04  & 0.1\\
0.2     & 0.06  & 0.07  & 0.1\\
0.3     & 0.06  & 0.09  & 0.1\\
0.4     & 0.06  & 0.09  & 0.1\\
0.5     & 0.06  & 0.09  & 0.1\\
0.6     & 0.04  & 0.09  & 0.1\\
0.7     & 0.02  & 0.08  & 0.1\\
0.8     & 0.014 & 0.09  & 0.1\\
0.9     & 0.01  & 0.07  & 0.07\\
\hline
\end{tabular}
\end{minipage}
\caption{{\small Maximal distances between the coefficients $C_{\pm}$ of (\ref{ae1}) and $s_{\kappa\lambda}$ of (\ref{specs}). Left: $\Delta(C_-,s)=\max_{\lambda}[s -C_-]$,  $ \Delta(s, C_+)=\max_{\lambda}[C_+ - s]$, and $\Delta(C_-, C_+)=\max_{\lambda}[C_+ - C_-]$
at different values of $\kappa$. Right: the same distances but with $\max_{\kappa}$ for different values of $\lambda$}.}
\end{table}

\subsection{Entanglement Entropy of Hawking Radiation}\label{ss:page}

The problem is as follows. Viewing black hole and its radiation as a
bipartite quantum system (\ref{SBE}) -- (\ref{stsp}), denote
\begin{equation}
\dim \mathcal{H}_{\mathfrak{B}}=\mathsf{L},\;\dim \mathcal{H}_{\mathfrak{E}}=%
\mathsf{K},\;\dim \mathcal{H}_{\mathfrak{S}}=\mathsf{KL}=\mathsf{N}.
\label{dims}
\end{equation}%
and index the bases in the state spaces $\mathcal{H}_{\mathfrak{B}}$ and $%
\mathcal{H}_{\mathfrak{E}}$ of its parties by $l=1,\dots ,\mathsf{L}$ and $%
k=1,\dots ,\mathsf{K}$. 

Note that in the discussed above case of free fermions, where the
description reduces to the one-body picture, the indexing sets of the block
and
its environment are $\Lambda $ and $\Omega \setminus \Lambda $,
but in that case $|\Omega |=|\Lambda |+|\Omega \setminus \Lambda |$, while
in (\ref{dims}) we have $\mathsf{N=KL}$. This is because the second
quantization is a kind of the "exponentiation" of the one-body picture.

Assuming the complete ignorance of the structure of the whole system $%
\mathfrak{S}$ (an evaporating black hole and its radiation), one can choose
as its ground state%
\begin{equation}
|\Psi _{\mathfrak{S}}\rangle=\{\Psi _{kl}\}_{k,l=1}^{\mathsf{K,L}}
\label{Psi}
\end{equation}%
the random vector uniformly distributed over the unit sphere in 
$\mathcal{H}_{\mathfrak{S}}=\mathcal{H}_{\mathfrak{B}} \otimes \mathcal{H}_{\mathfrak{E}}
\mathbb{=C}^{\mathsf{N}},$ $\mathsf{N=KL}$ (see (\ref{stsp}) and (\ref{dims}%
)). Thus, the density matrix $\rho _{\mathsf{N}}$ of $\mathfrak{S} $ and the
reduced density matrix $\rho _{\mathsf{LN}}$ of $\mathfrak{B}$ (the radiation)
 are
\begin{equation}
(\rho _{\mathsf{N}})_{k_{1}l_{1},k_{2}l_{2}}=\Psi _{k_{1}l_{1}}\overline{%
\Psi _{k_{2}l_{2}}},\;\;(\rho _{\mathsf{LN}})_{l_{1}l_{2}}=\sum_{k=1}^{%
\mathsf{K}}\Psi _{kl_{1}}\overline{\Psi _{kl_{2}}},  \label{rhopa}
\end{equation}%
(cf. (\ref{gss}) -- (\ref{rdmg})).

It is of interest to find the typical behavior of the corresponding (random)
entanglement entropy (see (\ref{ente}). It was suggested in
\cite{Pa:93a}, as a first step in this program, that%
\begin{equation}
\mathbf{E}\{S_{\mathsf{LN}}\}=\sum_{t=\mathsf{K}+1}^{%
\mathsf{KL}}\frac{1}{t}-\frac{\mathsf{L}-1}{2\mathsf{K}},\;\mathsf{L}%
\leq \mathsf{K}=\mathsf{N/L}.  \label{pag}
\end{equation}%
Formula (\ref{pag}) was then proved by using an explicit and rather involved
form of the joint eigenvalue distribution of random matrix $\rho_{\mathsf{LN}%
}$ (\ref{rhopa}), see e.g. \cite{Da-Co:14,Bi-Co:22} for reviews.


It follows from (\ref{pag}) that the two term asymptotic formula for large $%
\mathsf{K}$ and any $\mathsf{L}$, i.e., for%
\begin{equation}
1\lesssim \ \mathsf{L}\ll \ \mathsf{K}\ \lesssim \mathsf{N}  \label{bup0}
\end{equation}%
(cf. (\ref{bip})), is
\begin{equation}
\mathbf{E}\{S_{\mathsf{LN}}\}=\log \mathsf{L}-\frac{\mathsf{L}}{2\mathsf{K}}%
+O(1/\mathsf{K}),\;\mathsf{K}=\mathsf{N/L}\rightarrow \infty .  \label{paas}
\end{equation}%

Note that here we follow \cite{Pa:93a} and use the (cf. (\ref{bip})) standard natural $%
\log :=\ln $ to the base $e=$2.7182 instead $\log _{2}$ as in the definition
(\ref{ente}) of the von Neumann entropy.

It follows from (\ref{pag}) that in the asymptotic regime of the successive
limits (first $%
\mathsf{K}\rightarrow \infty $, 
then  $\mathsf{L}\rightarrow \infty $ (cf. (\ref{ar1}%
) and (\ref{bup0})), 
we have
\begin{equation}
\lim_{\mathsf{K}\rightarrow \infty }\mathbf{E}\{S_{\mathsf{LN}}\}:=\mathbf{E}%
\{S_{\mathsf{L}}\}=\log \mathsf{L}+o(1),\ \mathsf{L}\rightarrow \infty .
\label{subs}
\end{equation}%
Moreover, the same  holds in the asymptotic regime of the simultaneous
limits $\mathsf{K}\rightarrow \infty $, $\mathsf{L}\rightarrow \infty $,
provided that $\mathsf{L}/\mathsf{K}=o(1)$, i.e., $0 \leq \alpha < 1$
(cf. (\ref{diff})).

This case can be viewed as that describing the very initial stage of the
black hole radiation. On the other hand, in the asymptotic regime
(cf. (\ref{diff}) and (\ref{KLN}))
\begin{equation}  \label{lipa}
\mathsf{K}\rightarrow \infty , \; \mathsf{L}\rightarrow \infty, \; \mathsf{L}%
/\mathsf{K}\rightarrow \lambda >0,
\end{equation}
another possible implementation of the analog of the heuristic inequalities (%
\ref{bip}) (cf. (\ref{diff}) with $\alpha =1$), we have (cf. (\ref{sla}))%
\begin{equation}
\mathbf{E}\{S_{\mathsf{LN}}\}=\log \mathsf{L}- \left\{%
\begin{array}{cc}
\lambda/2, & 0\leq \lambda \leq 1, \\
1/2\lambda+\log \lambda , & \lambda \geq 1.%
\end{array}%
\right.  \label{mul}
\end{equation}%
This case corresponds to a later stage of the black hole radiation.


It can also be shown that the fluctuations of $S_{\mathsf{LN}}$ vanish for
large $\mathsf{K}$ and $\mathsf{L}$, see e.g. \cite{Da-Co:14}.

Basing on the formula (\ref{mul}), an interesting scenario of the black hole
evaporation was proposed in \cite{Pa:05,Pa:93a}, see also \cite%
{Al-Co:21} for a recent review.

\begin{figure}[ht]
\center{\includegraphics[width=10cm]{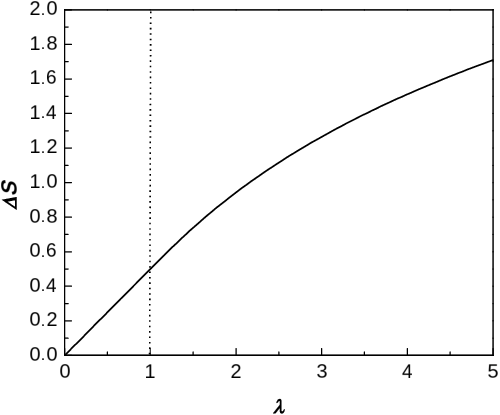}}
\caption{{\small The "deficit" $\Delta=-(S_{\mathsf{LN}} - \log \mathsf{L})$ of (\ref{mul}) (and (\ref{mulp1}) and (\ref{mulrr})) as function of $\lambda$.}}
\label{fig4}
\end{figure}

Here we only mention that function
given by the r.h.s. of (\ref{mul}) (see Figure \ref{fig4}) is monotone increasing, convex and piece-wise analytic. Its $p$th derivative has a jump from $0$ to $(-1)^{p}(p-1)!(p/2-1)$
for all $\;p\geq 3$,  "a phase transition" of the third order takes place.

Recall that the maximum of the von Neumann entropy (\ref{ente}) over the set
of $T\times T$ positive definite matrices of trace 1 is equal to $\log_{2}T$
. We conclude, following \cite{Pa:93a}, that while the (random) states (%
\ref{Psi}) (see also (\ref{PLO})) of the whole system are pure, the
subsystem states are typically quite close to the maximally mixed states
with the "deficit" given by the second term of the r.h.s. of (\ref{mul}).

The link of the above results with those of the previous subsection is as follows.
It was mentioned there that in
the case of free fermions the dimension $\dim \mathcal{H}_{\mathfrak{S}}$ of
the state space of the system $\mathfrak{S}$ and the volume $|\Omega
|=|N|^{d}$ of the domain occupied by $\mathfrak{S}$ are related as $|\Omega
|=|N|^{d}=\log _{2}\dim \mathcal{H}_{\mathfrak{S}}$ and the same for its
party $\mathfrak{B} $ occupying a subdomain $\Lambda :L^{d}=|\Lambda |=\log
_{2}\dim \mathcal{H}_{\mathfrak{B}} $. In fact, this logarithmic dependence
is general for the many-body quantum systems. Thus, viewing the black hole
and its radiation as the parties of a many-body bipartite system (see (\ref%
{SBE})) and taking into account (\ref{dims}), we can interpret $\log L$ in
the asymptotic formulas (\ref{subs}) -- (\ref{mul}) as the "volume" of the
spatial domain occupied by the black hole radiation, hence, these asymptotic
formulas are the analogs of the volume law (\ref{vol}), see (\ref{lbrmt}%
) and (\ref{sl1}) in particular.

We will show now
that the standard facts of random
matrix theory, that date back to the 1960s, provide a streamlined proof
of the validity of (\ref{subs}) -- (\ref{mul}) for a rather wide class of
random vectors including those of (\ref{Psi}) and not only for the expectation of the
entanglement entropy but also for its all typical realization, i.e., with
probability 1. One can say that these results manifests the typicality and
the universality of Page's formula, given by the r.h.s. of (\ref{mul}) and
(\ref{mulp1}). For other versions of these important properties see \cite{Da-Co:14,Bi-Co:22,Na-Co:18}.

Let
\begin{equation}
\{X_{lk}\}_{l,k=1}^{\infty },\;\mathbf{E}\{X_{lk}\}=\mathbf{E}%
\{X_{lk}^{2}\}=0,\;\mathbf{E}\{|X_{lk}|^{2}\}=\xi^2>0  \label{xjk}
\end{equation}%
be an infinite collection of independent identically distributed (i.i.d.)
complex random variables with zero mean and unit variance,%
\begin{equation}
X_{\mathsf{L} \mathsf{N} }=\{X_{lk}\}_{l,k=1}^{\mathsf{L},\mathsf{K}}
\label{XLO}
\end{equation}%
be the $\mathsf{K}\times \mathsf{L}$ matrix and%
\begin{equation}
Z_{\mathsf{L} \mathsf{N} }=\mathrm{Tr}X_{\mathsf{L} \mathsf{N} }X_{\mathsf{L}
\mathsf{N} }^{\ast }=\sum_{l=1}^{\mathsf{L}}\sum_{k=1}^{\mathsf{K}%
}|X_{lk}|^{2}.  \label{ZLO}
\end{equation}%
View $X_{\mathsf{L} \mathsf{N} }$ as a random vector in
\begin{equation}
\mathbb{C}^{\mathsf{N}}=\mathbb{C}^{\mathsf{K}}\otimes \mathbb{C}^{\mathsf{L}%
},\;\mathsf{N}=\mathsf{KL,}  \label{HS}
\end{equation}%
and (\ref{ZLO}) as the square of its Euclidian norm and introduce the corresponding
random vector of unit norm (cf. (\ref{Psi}))
\begin{equation}
\Psi_{\mathsf{N}} =X_{\mathsf{L} \mathsf{N} }/Z_{\mathsf{L} \mathsf{N}
}^{1/2}.  \label{PLO}
\end{equation}%
Note that if $\{X_{kl}\}_{k,l=1}^{\infty }$ are the complex Gaussian
random variables with zero mean and unit variance, then $\Psi _{\mathsf{N}}$ of (\ref{PLO}) is
uniformly distributed over the unit sphere of $\mathbb{C}^{\mathsf{N}}$ (see
(\ref{HS})), hence, coincides with (\ref{Psi}) and the setting of \cite{Pa:93a}.

\begin{result}
\label{r:page}
Consider a bipartite quantum systems having the random vector (\ref{PLO}) as its ground state.
Let $S_{\mathsf{LN}}$ be the entanglement entropy defined by (\ref{ente}), and (\ref{rhopa}) with $\Psi_{\mathsf{N}}$ of (\ref{PLO}). Then we have the analogs of (\ref{subs}) -- (\ref{mul}) valid for all typical realizations of (\ref{PLO}) (with probability 1 with respect to (\ref{xjk}))
\begin{equation}
\lim_{\mathsf{K}\rightarrow \infty } S_{\mathsf{LN}}:=S_{\mathsf{L}}=\log \mathsf{L}+o(1),\ \mathsf{L}\rightarrow \infty,
\label{subsp1}
\end{equation}
and
\begin{equation}\label{mulp1}
S_{\mathsf{LN}}=\log \mathsf{L}- \left\{%
\begin{array}{cc}
\lambda/2, & 0\leq \lambda \leq 1, \\
1/2\lambda+\log \lambda, & \lambda \geq 1,%
\end{array}%
\right.+o(1),
\;\;  \mathsf{K,L}\rightarrow \infty, \; \mathsf{L}%
/\mathsf{K}\rightarrow \lambda >0.
\end{equation}
\end{result}

One can say that these results manifests the typicality (the validity with probability 1) and
the universality (the independence of the probability law of $\{X_{jk}\}$) in (\ref{xjk})) of Page's formula, given by the r.h.s. of (\ref{mul}) and
(\ref{mulp1}). For other versions of these important properties see \cite{Da-Co:14,Bi-Co:22,Na-Co:18}.

The proof of the result is given in Appendix \ref{a:page}.

\section{Conclusions}
Our main motivation was to study possible asymptotic forms of entanglement entropy of quantum bipartite systems in a regime where the size of one of the parties (block) grows simultaneously with the size of the system. We believe that this regime is of interest both in itself and because it seems more adequate for interpreting numerical results. The regime can be considered for various cases of interaction radii and hopping in the Hamiltonian of the system. 

Using a random matrix as a one-body Hamiltonian can serve as a model for long-range hopping whose radius is of the same order of magnitude as the size of the system. 
We show that in this case the asymptotic behavior of the entanglement entropy follows the volume law, but not the area law or the enhanced area law, that arises in the case of finite-range hopping and the widely used asymptotic regime in which the block size is considered large only after a macroscopic limit passage for the entire system. 

For the proof, we use both new seemingly quite general two-sided bounds for the entanglement entropy and existing rigorous results from random matrix theory. The latter also proved to be useful for analyzing the generalization of the Hawking radiation model in the theory of black holes. This analysis, which turns out to be fairly simple and transparent, is also presented in the paper. It implies the validity of the Page formula, obtained initially for a particular case, in quite wide class of typical random states of the system.

\vspace{6pt}

\section{Acknowledgments}
L.P. is grateful to Ecole Normale Superieure (Paris) and l'Institut des Hautes Etudes Scientifiques (Bures-sur-Ivettes) for their kind hospitality during the first stage of the work. Special thanks are due to Prof. E. Brezin for many interesting discussions. V.S. acknowledges the support from the Project IMPRESS-U: N2401227.

\appendix

\section{Proof of Result \protect\ref{r:1}.}\label{a:hbou}
The proof is based on the following properties of $h$ and $h_0$ of (\ref{h}) and (\ref{h0h}) which are easy to check, see, e.g. \cite{El-Co:16}:

\medskip
(i) $h$ is concave ($h''(x) \le 0, \; x \in (0,1)$) and
\begin{equation}\label{hlb}
h(x)\ge 4x(1-x), \; x \in [0,1];
\end{equation}

\medskip

(ii) $h_0$ of (\ref{h0h})
is also concave ($h_0''(y) \le 0, \; y \in (0,1/4)$) ) and
\begin{equation}\label{h4h0}
h_0(y)=-y\log_2 y +O(y), \; y \searrow 0.
\end{equation}
To get the lower bound in (\ref{tsb}), we denote by $\{p _{\alpha
}\}_{\alpha =1}^{|\Lambda|}$ the eigenvalues of $P _{LN }$ of (\ref{fpr}) and use (\ref{hlb}) and (\ref{lub}):
\begin{align}
&\hspace{-1cm}S_{LN }=\Tr h(P_{LN })= \sum_{\alpha=1 }^{|\Lambda|}h(p _{\alpha }) \notag
\\&\ge \sum_{\alpha=1 }^{|\Lambda|}4p _{\alpha }(1-p_\alpha)
=4\Tr P _{LN }(\mathbf{1} _{\Lambda}-P _{LN})
=:\mathcal{L}_{LN}.
\label{lb}
\end{align}%
To get the upper bound in (\ref{tsb}), we use (\ref{lub}) (or (\ref{lb})), (\ref{h0h}), and the concavity of
$h_0$, implying by the Jensen inequality
\begin{align*}
&\hspace{-1cm}S_{LN} =\sum_{\alpha=1 }^{|\Lambda |} h (p_\alpha )=\sum_{\alpha=1 }^{|\Lambda |} h_{0}(p _{\alpha} (1-p_\alpha) ) \\
&=|\Lambda |\;\Big( |\Lambda |^{-1}\sum_{\alpha=1}^{|\Lambda
|}h_{0}(p_\alpha (1-p_\alpha)\Big) \leq |\Lambda |\;h_{0}\Big( (4|\Lambda
|)^{-1}\;
\mathcal{L}_{\Lambda\Omega})=:\mathcal{U}_{LN }.
\end{align*}%
To get (\ref{tsbr}) -- (\ref{lubr}), we just apply the expectation to (\ref{tsb}) and use once more the Jensen inequality in the r.h.s.\begin{equation*}
\mathbf{E}\{\mathcal{U}_{LN}\}=|\Lambda|\mathbf{E}\{h_0((4|\Lambda)^{-1}|\mathcal{L}_{LN}\} \le|\Lambda|h_0\big(
(4|\Lambda)^{-1}|\mathbf{E}\{\mathcal{L}_{LN}\}\big).
\end{equation*}

\section{Proof of Result \protect \ref{r:rmlob}}
\label{a:m4}

We will use the bounds  (\ref{tsb}) -- (\ref{h0h}). It follows
from (\ref{pdma}) that the expectation of the lower bound $\mathcal{L}%
_{LN}$ is expressed via the mixed fourth moments of the entries $%
\{U_{kl}\}_{k,l=1}^{N}$ of the Haar distributed unitary matrix $U_{N}$. The
moments are known (see, e.g. \cite{Pa-Sh:11}, Problem 8.5.2)
 \begin{align}
&\hspace{-1cm} \mathbf{E}\{U_{a_{1}b_{1}}%
\overline{U}_{\alpha _{1}\beta
_{1}}U_{a_{2}b_{2}}\overline{U}_{\alpha
_{2}\beta _{2}}\} \notag \\
& =(n^{2}-1)^{-1}\big(\delta _{a_{1}\alpha _{1}}\delta _{b_{1}\beta
_{1}}\delta _{a_{2}\alpha _{2}}\delta _{b_{2}\beta _{2}}+\delta
_{a_{1}\alpha _{2}}\delta _{b_{1}\beta _{2}}\delta _{a_{2}\alpha _{1}}\delta
_{b_{2}\beta _{1}}\big)\notag \\
& -(n(n^{2}-1))^{-1}\big(\delta _{a_{1}\alpha _{2}}\delta _{b_{1}\beta
_{1}}\delta _{a_{2}\alpha _{1}}\delta _{b_{2}\beta _{2}}+\delta
_{a_{1}\alpha _{1}}\delta _{b_{1}\beta _{2}}\delta _{a_{2}\alpha _{2}}\delta
_{b_{2}\beta _{1}}\big)
\label{ums}
\end{align}
and we obtain
for $1\leq l<m\leq N$%
\begin{equation*}
\mathbf{E}\{|P_{lm}|^{2}\}=\sum_{k_{1},k_{2}=1}^{K}\mathbf{E}%
\{U_{lk_{1}}U_{mk_{2}}\overline{U_{mk_{1}}}\, \overline{U_{lk_{2}}}\}=\frac{1%
}{ (N^{2}-1)}\sum_{k_{1},k_{2}=1}^{K}(\delta_{k_{1}k_{2}}-N^{-1})\frac{K(N-K)%
}{N(N^{2}-1)}.
\end{equation*}%
This and (\ref{lub}) imply
\begin{equation}  \label{lubm}
\mathbf{E}\{\mathcal{L}_{LN }\}=4\sum_{l=1}^{L}\sum_{m=L+1}^N
\mathbf{E}\{|P_{lm}|^{2}\}=4\frac{K(N-K)L(N-L)}{N(N^{2}-1)},
\end{equation}%
and then (\ref{tsb}) and (\ref{Mma}) yield (\ref{lbrmt}).

\section{Proof of Result \protect\ref{r:rmas}}

\label{a:rmha}

We will use formulas (\ref{dosp}) -- (\ref{slgen}) that reduce the problem of
the asymptotic analysis of the entanglement entropy to that of the Counting
Measure $\mathcal{N}_{P_{_{LN }}}$ (\ref{dosp}) of the random
matrix $P_{LN }$ (\ref{pdlma}) in the regime (\ref{kln}), more
precisely, the limit
\begin{equation}
\nu _{P}=\lim \mathcal{\ N}_{P_{LN }}/L  \label{ncmp}
\end{equation}%
in the sense of (\ref{kln}) (see also Appendix \ref{a:page} for a similar
approach).

The explicit form of $\nu_P$ has actually been known since 1980 and
is called the Wachter distribution. It was obtained in \cite{Wa:78} in the
context of statistics, and according to this work, convergence in (\ref{ncmp})
is in probability. In the subsequent works \cite%
{Pa-Sh:11,Mi-Sp:17,Pa:22,Va:01} the distribution was obtained by other
methods and in other settings, in particular, the convergence with
probability 1 was also proved.

We have, according to these works
\begin{align}
&\hspace{1.5cm}\nu _{P} =m_{0}\delta _{0}+m_{1}\delta _{1}+\nu _{ac},
\label{wach} \\
m_{0} &=\max (\lambda -\kappa, 0)/\lambda ,\;m_{1}=\max (\lambda +\kappa-1, 0)/\lambda ,  \notag
\end{align}%
and the density $\nu _{ac}^{\prime }$ of $\nu _{ac}$ in (\ref{wach}) is
given by (\ref{rho}). Now, plugging $\nu_P$ into the divided   by $L$ version of (\ref{slgen}) for our case, and assuming (%
\ref{kln}), we obtain (\ref{sl1}) -- (\ref{rho}), taking into account that
the atoms in (\ref{wach}) do not contribute to the integral in the r.h.s. of the limiting
form of (\ref{slgen}) because of equalities $h(0)=h(1)=0$.


Note that the atom at 0 in (\ref{wach}) can be obtained just by calculating
the rank of (\ref{pdlma}) (cf. (\ref{mop})) and the atom at 1 has the same
origin, because we have in view of the unitarity of $U_N$:
\begin{equation}
(P_{LN})_{l_{1}l_{2}}=\sum_{k=1}^K U_{l_1 k}\overline{U_{kl_2}}=\delta_{l_1 l_2}-\sum_{k=K+1}^N U_{l_1 k}\overline{U_{kl_2}}.  \label{ploo}
\end{equation}



\section{Proof of (\protect\ref{eel1})}

\label{a:L1}

It follows from (\ref{SBD}), (\ref{h}), and  (\ref{h0h}) that (see also (\ref{frlat}))
\begin{equation}\label{sh0}
S_{LN}=\Tr h_0(P_{LN}(\mathbf{1}_\Lambda-P_{LN})).
\end{equation}
It is easy to see that if $L=N-1$, then %
\begin{equation*}
(P_{LN}(\mathbf{1}_\Lambda-P_{LN}))_{l_1l_2}=(P_{\Omega})_{l_{1}N}\overline{%
(P_\Omega)_{l_{2}N}},\;l_{1},l_{2}=1,\ldots ,(N-1),
\end{equation*}
i.e., it is a hermitian operator of rank one. Hence its eigenvalues are 0 of
multiplicity $(L-1)$ and
\begin{equation*}
q_{N}=\sum_{l=1}^{N-1}|(P_{\Omega})_{lN}|^2  \label{pcon}
\end{equation*}
of multiplicity 1. This, (\ref{pdma}), and the unitarity of $U_N$ imply
\begin{equation}
q_{N}=u_{N}(1-u_{N}), \; u_{N}= \sum_{l=1}^{N-1}|U_{lN}|^2,  \label{qul1}
\end{equation}
and  then, by (\ref{sh0})
\begin{equation*}
S_{LN}|_{L=N-1}=h_0(u_{N}(1-u_{N})),  \label{eel11}
\end{equation*}
It follows then from the explicit form of the forth mixed moments of the
entries $\{U_{lk}\}_{l,k=1}^{N}$ of the Haar distributed matrix $U_{N}$ (see (\ref{ums})) that
\begin{align*}
&\hspace{2cm}\overline{ u}_{N}:=\mathbf{E}\{u_{N}\}=K/N= \kappa(1+O(1/N)), \\
&\mathbf{Var}\{u_{N}\}:==\mathbf{E} \{|u_{N}-\overline{ u}_{N}|^{2}\}\mathbf{=E} \{|u-K/N|^{2}\}\leq \kappa(1-\kappa)N^{-1}(1+O(1/N)),\;N\rightarrow \infty.  \notag  \label{uev}
\end{align*}%
Hence, by the Tchebychev inequality, we have for any $\varepsilon >0$%
\begin{equation}
\mathbf{P}\{|u_{N}-K/N|>\varepsilon \} \leq \mathbf{Var}\{u_{N}\}/\varepsilon^2 \leq
\kappa(1-\kappa)/\varepsilon ^{2} N.  \label{ps1e}
\end{equation}%
We conclude by the continuity of $h_0$ that for the overwhelming majority of
realizations (with probability $1-\varepsilon $ for any $\varepsilon >0$) we have%
\begin{equation}
S_{LN}|_{L=N-1}=h_{0}(\kappa(1-\kappa ))+o(1 ), \; \varepsilon \to 0.\;
\label{s1e}
\end{equation}%
For instance, we can choose $\varepsilon =N^{-1/3}$ to have $O(1/N^{1/3})$
in the r.h.s. of (\ref{ps1e}) and (\ref{s1e}).

\section{Proof of Result \protect\ref{r:page}}
\label{a:page}

Following (\ref{Psi}) --  \ref{rhopa}) and (\ref{xjk}) -- (\ref{PLO}), we obtain for the corresponding reduced density matrix
\begin{equation}
\rho _{\mathsf{L}\mathsf{N}}=X_{\mathsf{L}\mathsf{N}}X_{\mathsf{L}\mathsf{N}%
}^{\ast }/Z_{\mathsf{L}\mathsf{N}}.  \label{rdmrs}
\end{equation}%
Dividing the numerator and denominator by $\xi^2 K$, we obtain%
\begin{equation}
\rho _{\mathsf{L}\mathsf{N}}=W_{\mathsf{L}\mathsf{N}}/
\mathcal{Y}_{\mathsf{L}\mathsf{N}%
}.  \label{ryw}
\end{equation}%
where%
\begin{align}
&W_{\mathsf{L}\mathsf{N}}=\mathcal{ X}_{\mathsf{L}\mathsf{N}}^{\ast }\mathcal{X}_{\mathsf{L}%
\mathsf{N}}/\mathsf{K},\ \mathcal{Y}_{\mathsf{L}\mathsf{N}}=\mathcal{Z}_{\mathsf{L}\mathsf{N}}/%
\mathsf{K}, \;
\mathcal{Z}=\sum_{l,k=1}^{L,K}|\mathcal{X}_{lk}|^2, \label{yw}
\end{align}%
and $\{\mathcal{X}_{lk}\}_{l,k=1}^\infty$ are independent identically distributed random variables such that (cf. (\ref{xjk}))
\begin{equation}\label{cxjk}
\mathbf{E}\{\mathcal{X}_{lk}\}=\mathbf{E}\{\mathcal{X}^2_{lk}\}=0, \; \mathbf{E}\{|\mathcal{X}_{lk}|^2\}=1.
\end{equation}
This allows us to write the von Neumann entropy of (\ref{ryw}) as%
\begin{align}
& S_{\mathsf{L}\mathsf{N}}=\log \mathcal{Y}_{\mathsf{L}\mathsf{N}}-\mathcal{Y}_{\mathsf{L}%
\mathsf{N}}^{-1}\Tr W_{\mathsf{L}\mathsf{N}}\log W_{\mathsf{L}\mathsf{N}},
\notag \\
& \hspace{1cm}=\log -\mathcal{Y}_{\mathsf{L}\mathsf{N}}--\mathcal{Y}_{\mathsf{L}\mathsf{N}%
}^{-1}\int w\log w\,\nu _{W_{\mathsf{L}\mathsf{N}}}(dw),  \label{sdos}
\end{align}%
where
\begin{equation}
\nu _{W_{\mathsf{L}\mathsf{N}}}=\mathcal{N}_{W_{\mathsf{L}\mathsf{N}}}/%
\mathsf{L}  \label{ncmw}
\end{equation}%
is the Normalized Counting Measure of eigenvalues of $W_{\mathsf{L}\mathsf{N}%
}$ (cf. (\ref{dosp})).

It follows from the Strong Law of Large Numbers for the collection (\ref{xjk}%
) that we have with probability 1 for any $\mathsf{L}$ and $\mathsf{K}
\rightarrow \infty $
\begin{equation}
\mathcal{Z}_{\mathsf{L} \mathsf{N} }=\sum_{k,l=1}^{\mathsf{K,L}}|\mathcal{X}_{kl}|^{2}=\mathsf{KL%
}(1+o(1)),\;\mathsf{N:=KL}\rightarrow \infty.  \label{zlln}
\end{equation}%
Thus, the first term on the right of (\ref{sdos}) is in view of (\ref{yw})
\begin{equation}\label{lyln}
\log \mathcal{Y}_{\mathsf{L} \mathsf{N} }=\log \mathsf{L}+o(1),\;\mathsf{K}\rightarrow
\infty ,
\end{equation}%
i.e., it coincides with that in (\ref{paas}). Note that this is valid in the
both asymptotic regimes, the subsequent limits $\ \mathsf{K}\rightarrow
\infty $ and then $\mathsf{L}\rightarrow \infty $ and the simultaneous
limits
(\ref{lipa})
and for all typical realizations (with probability 1).

Consider now the second term on the right of (\ref{sdos}) and assume first
that $\mathsf{L}$ is fixed while $\mathsf{K}\rightarrow \infty $. Since (see
(\ref{yw}))%
\begin{equation}
(W_{\mathsf{L}\mathsf{N}})_{l_{1}l_{2}}=K^{-1}\sum_{k=1}^{\mathsf{K}%
}\mathcal{X}_{l_{1}k}\mathcal{X}_{l_{2}k}^{\ast },\;l_{1},l_{2}=1,\ldots ,\mathsf{K},
\label{ropa}
\end{equation}%
it follows, again from the Strong Law of Large Numbers and (\ref{xjk}) that $%
W_{\mathsf{L}\mathsf{N}}$ converges with probability 1 as $\mathsf{K}%
\rightarrow \infty $ to the $\mathsf{L}\times \mathsf{L}$ unit matrix $%
\mathbf{1}_{\mathsf{L}}$, thus the second term on the right of (\ref{sdos})
vanishes as $\mathsf{K}\rightarrow \infty $. We conclude that in the regime
of successive limits (see (\ref{lipa})) we have with probability 1%
\begin{equation}
S_{\mathsf{L}}:=\lim_{\mathsf{K}\rightarrow \infty }S_{\mathsf{L}\mathsf{N}%
}=\log \mathsf{L},  \label{subs1}
\end{equation}%
i.e., the version of (\ref{subs}) but now for all typical realizations and
for not necessarily Gaussian $X_{\mathsf{L}\mathsf{N}}$ of (\ref{XLO}) with i.i.d. components (\ref%
{xjk}). In the context of black hole model of \cite{Pa:93a,Pa:05}
this result corresponds to the maximum mixed state of the Hawking
radiation despite the pure   initial state (or (\ref{Psi}) of the whole system
(black hole and radiation).

Passing to the regime of the simultaneous limits (\ref{lipa}), we note first that the matrix $W_{\mathsf{L}%
\mathsf{N}}$ is known in statistics since the late 1920s as the sample
covariance matrix of the sample of $\mathsf{K}$ random $\mathsf{L}$%
-dimensional vectors $\mathcal{X}_{k}=\{\mathcal{X}_{lk}\}_{l=1}^{\mathsf{L}},\;k=1,\dots,%
\mathsf{K}$ (data vectors) and according to \cite{Ma-Pa:67} (see also \cite%
{Pa-Sh:11} for a review) the Normalized Counting Measure $\nu _{W_{\mathsf{L}%
\mathsf{N}}}$ (\ref{ncmw}) of eigenvalues of $W_{\mathsf{L}\mathsf{N}}$
converges with probability 1 to the non random limit (cf. (\ref{wach}))%
\begin{align}
& \lim_{\mathsf{K\rightarrow \infty ,L\rightarrow \infty ,L/K}\rightarrow
\lambda}\nu _{W_{\mathsf{L}\mathsf{N}}}=:\nu _{W}=\max \{1-\lambda,0\}\delta
_{0}+\nu _{ac},  \label{mp1} \\
& \nu _{ac}^{\prime }(w)=\frac{\sqrt{(w_{+}-w)(w-w_{-})}}{2\pi \lambda w}%
\mathbf{1}_{[w_{-},w_{+}]},\;w_{\pm }=(1\pm \sqrt{\lambda })^{2}.  \notag
\end{align}%
Note that the atom at 0 in (\ref{mp1}) is just (cf. (\ref{ploo}))
\begin{equation}
\lim_{\mathsf{K\rightarrow \infty ,L\rightarrow \infty ,L/K}\rightarrow
\lambda} (\mathsf{L} - \mathrm{rank} \, W_{\mathsf{LN}})/\mathsf{L}.
\label{mop}
\end{equation}
It follows from (\ref{ncmw}) -- (\ref{lyln}), and  (\ref{mp1}) that the simultaneous
limit (\ref{lipa}) of the second term in  (\ref{sdos}) is
\begin{equation}\label{I}
I=\frac{1}{2\pi \lambda }\int_{w_{-}}^{w_{+}}\sqrt{(w-w_{+})(w-w_{-})}\,\log
wdw.
\end{equation}%
We will use the identity%
\begin{equation*}
\log w=\int_{0}^{\infty }\left( \frac{1}{t+1}-\frac{1}{t+w}\right) dt,
\end{equation*}%
implying
\begin{equation}\label{I1-I2}
I=\lim_{A\rightarrow \infty }(I_{1}-I_{2}),
\end{equation}%
where%
\begin{eqnarray}
I_{1} &=&\frac{1}{2\pi \lambda }\int_{0}^{A}\frac{dt}{t+1}%
\int_{w_{-}}^{w_{+}}\sqrt{(w-w_{+})(w-w_{-})}dw,\; \notag \\
I_{2} &=&\frac{1}{2\pi \lambda }\int_{0}^{A}dt\int_{w_{-}}^{w_{+}}\frac{%
\sqrt{(w-w_{+})(w-w_{-})}}{t+w}dw.
\label{I1I2}
\end{eqnarray}%
Changing variables in the integral over $w$  in $I_1$ to
\begin{equation}
w=m+lx,\;m=\frac{w_{+}+w_{-}}{2}=(1+\lambda ),\;\frac{w_{+}-w_{-}}{2}=2\sqrt{%
\lambda },  \label{cvar}
\end{equation}%
we obtain%
\begin{equation}
I_{1}=\frac{2}{\pi }\int_{0}^{A}\frac{dt}{t+1}\int_{-1}^{1}\sqrt{1-x^{2}}%
dx=\int_{0}^{A}\frac{dt}{t+1}=\log A+O(1),\;A\rightarrow \infty .  \label{i1}
\end{equation}%
The same change of variables (\ref{cvar}) yields  %
\begin{equation*}
I_{2}=\frac{l}{2\pi \lambda }\int_{0}^{A}dt\int_{-1}^{1}\frac{\sqrt{1-x^{2}}%
}{x+\tau }dx,\;\tau =\frac{t+m}{l}>1.
\end{equation*}%
The integral over $x$ is $\pi (\tau - (\tau^2 -1)) $, hence, by (\ref{cvar})%
\begin{eqnarray}
I_{2} &=&2\int_{m/l}^{(A+m)/l}\left( \tau -(\tau ^{2}-1)^{1/2}\right) d\tau
\notag \\
&=&\left. \left( \frac{1}{2}\log (\tau +(\tau ^{2}-1)^{1/2})-\frac{1}{4}%
(\tau -(\tau ^{2}-1)^{1/2})^{2}\right) \right\vert _{\tau =m/l}^{(A+m)/l},
\label{mim}
\end{eqnarray}%
Now, taking into account (\ref{cvar}),
\begin{eqnarray}
\left. \left( \tau +(\tau ^{2}-1)^{1/2}\right) \right\vert _{\tau =m/l}
&=&2l^{-1}\max (1,\lambda ),\;\left. \left( \tau -(\tau ^{2}-1)^{1/2}\right)
\right\vert _{\tau =m/l}=\notag\\
& & =2l^{-1}\min (1,\lambda ),  \notag \\
\left. \left( \tau +(\tau ^{2}-1)^{1/2}\right) \right\vert _{\tau =(A+m)/l}
&=&2Al^{-1}+O(1/A),\;A\rightarrow \infty ,  \label{mmax}
\end{eqnarray}%
and  (\ref{I1-I2}) -- (\ref{mmax}), we get for $I$ of (\ref{I})
\begin{equation}\label{mulrr}
I=2^{-1}\min(\lambda,\lambda^{-1})+\log \; \max(1,\lambda).
\end{equation}
This coincides with the second term of the r.h.s. of (\ref{mul}), hence proves  with probability 1 Result \ref{r:page}.



\end{document}